\documentstyle[epsfig,12pt,dina4]{article}
\input{epsf}                    

\begin{document}

\baselineskip=18pt     

\noindent
{\Large THERMODYNAMIC VARIABLES FROM SPECTATOR DECAY\footnote{
presented at 14th Winter Workshop on Nuclear Dynamics, Snowbird, Utah,
31 Jan - 7  Feb 1998}
}
\vspace{0.6cm}

\baselineskip=14pt
\noindent
\hspace*{2.0cm}W. Trautmann$^1$ for the ALADIN collaboration$^2$
\vspace{0.2cm}

\noindent
\hspace*{2.0cm}$^1$Gesellschaft  f\"ur  
Schwerionenforschung mbH\\
\hspace*{2.0cm}D-64291 Darmstadt\\
\hspace*{2.0cm}Germany\\
\hspace*{2.0cm}$^2$Catania, Darmstadt, East Lansing, Frankfurt,\\
\hspace*{2.0cm}Milano, Moscow, Rossendorf, Warsaw
\vspace{1.0cm}

\baselineskip=18pt

\noindent
{\large\bf INTRODUCTION}
\vspace{0.3cm}

The Van-der-Waals-type range dependence of the nuclear forces has provided a 
good part of the motivation for fragmentation studies. The predictions 
of a liquid-gas phase transition in nuclear 
matter \cite{jaqa83,scott} have 
raised the hope that signals of it may be identified 
in reactions of finite nuclei.
The observation of multifragmentation \cite{jakob,fried,moretto} 
represented a major breakthrough in this direction, as it indicated that
the created nuclear systems may actually pass through states of high
temperature and low density during the later stages of the reaction.
Multifragmentation has been predicted \cite{gross90,bond95}
to be the dominant decay mode under such conditions that are similar to
those expected for the coexistence region in the 
nuclear-matter phase diagram.

Many features of multifragmentation are well reproduced by the statistical 
multifragmentation models \cite{gross90,bond95,botv95}.
They predominantly include the fragment distributions and correlations
describing the populated partition space. But also kinetic observables such 
as the energies or velocity correlations of the produced fragments have 
been found to agree with the statistical predictions in certain 
cases \cite{oesch,kwiat}. The essential assumption 
on which these models are based is that of a single equilibrated breakup
state at the end of the dynamical evolution and of a statistical 
population of the corresponding phase space. To the extent that it is
realized in nature, this scenario offers the possibility to map out the
nuclear phase diagram, even though for finite nuclei, by sampling
the thermodynamic equilibrium conditions associated with the 
multi-fragment breakup channels. 

Considerable progress has been made with this program during recent years. 
Multifragmentation has been studied over a wide range of different classes
of high-energy reactions and breakup temperatures have been deduced from
the measured data. 
The correlation of
the temperature with the excitation energy, often referred to as 
the caloric curve of nuclei and first reported for spectator decays
following $^{197}$Au + $^{197}$Au reactions at 600 MeV per 
nucleon \cite{poch95}, has also been derived for other 
cases \cite{kwiat,haug96,ma97}. Critical evaluations of the 
obtained results and of the applied methods have followed.
More work is clearly needed in order to complete the 
picture \cite{poch97}.

In this contribution, the focus will be on the decay of excited spectator 
nuclei \cite{dronten}. New results will be reported 
that were obtained in
two recent experiments with the ALADIN spectrometer at SIS in which
reactions of $^{197}$Au on $^{197}$Au in the regime of 
relativistic energies up to 1 GeV per nucleon were studied.
In the first experiment, the ALADIN spectrometer was used
to detect and identify the products of the projectile-spectator
decay \cite{schuetti}. The Large-Area Neutron Detector (LAND)
was used do measure coincident free neutrons emitted by the 
projectile source. In the second experiment, three multi-detector
hodoscopes, consisting of a total of 216 Si-CsI(Tl) telescopes,
and three high-resolution telescopes were positioned at backward angles
to measure the yields and correlations of isotopically resolved
light fragments of the target-spectator decay \cite{hongfei}.
From these data excitation energies and masses, temperatures,
and densities were deduced. Before proceeding to the discussion
of these new data some of the indications
for equilibration during the spectator decay will be briefly recalled. 
\vspace{0.8cm}

\noindent
{\large\bf EVIDENCE FOR EQUILIBRATION}
\vspace{0.3cm}

The universal features of the spectator decay,
as apparent in the
observed $Z_{bound}$ scaling of the measured charge 
correlations, were the first and perhaps most striking indications 
for equilibrium \cite{schuetti}.
The quantity $Z_{bound}$ is defined
as the sum of the atomic numbers $Z_i$ of all projectile fragments
with $Z_i \geq$ 2. It represents the charge of the
original spectator system
reduced by the number of hydrogen isotopes emitted during its decay.

\begin{figure}[ttb]
   \centerline{\epsfig{file=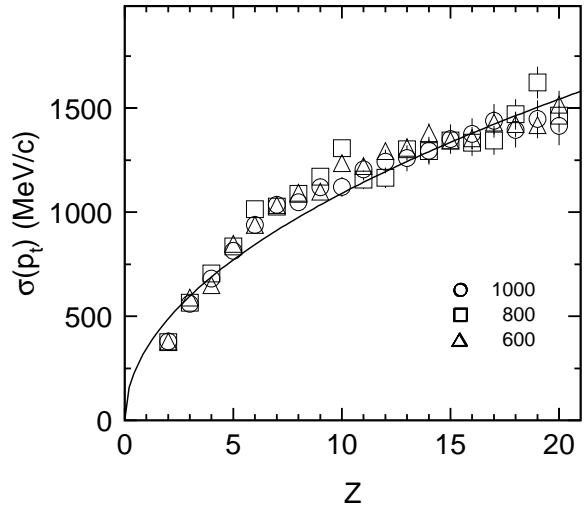,height=7cm}}
\caption[]{\it\small
Widths of the transverse-momentum distributions
$\sigma (p_t)$ as a function of the fragment atomic number $Z$ 
for the reactions  $^{197}$Au on $^{197}$Au at
$E/A$ = 600, 800, and 1000 MeV and for 20 $\le Z_{bound} \le$ 60.
The line is proportional to $\sqrt{Z}$ (from Ref. \cite{schuetti}).
}
\end{figure}

The invariance of the fragmentation patterns, when plotted as a function
of $Z_{bound}$, suggests that the memory of the entrance channel 
and of the dynamics governing the primary interaction of the 
colliding nuclei is lost.
This feature extends to other observables. The transverse-momentum
widths of the fragments, as shown in Fig. 1, 
do not change with the bombarding energy,
indicating that collective contributions to the transverse motion are 
small. 
The equilibration of the three kinetic degrees of freedom in 
the moving frame of the projectile spectator was confirmed by the 
analysis of the measured velocity spectra \cite{schuetti}.
The square-root dependence on the atomic number $Z$ 
implies kinetic energies nearly independent of $Z$ and hence of the mass.

\begin{figure}[ttb]
   \centerline{\epsfig{file=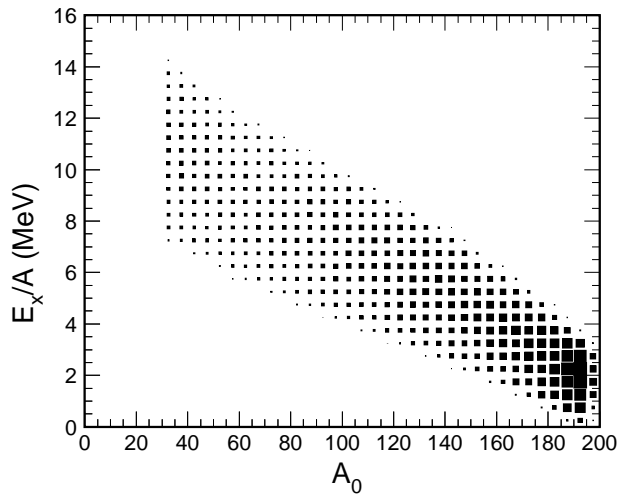,height=7cm}}
\caption[]{\it\small
Excitation energy $E_x/A$ as a function of the mass $A_0$ for the
ensemble of excited spectator nuclei
used as input for the calculations with the
statistical multifragmentation model. The area of the squares
is proportional to the intensity (from Ref. \cite{hongfei}).
}
\end{figure}

\begin{figure}[ttb]
   \centerline{\epsfig{file=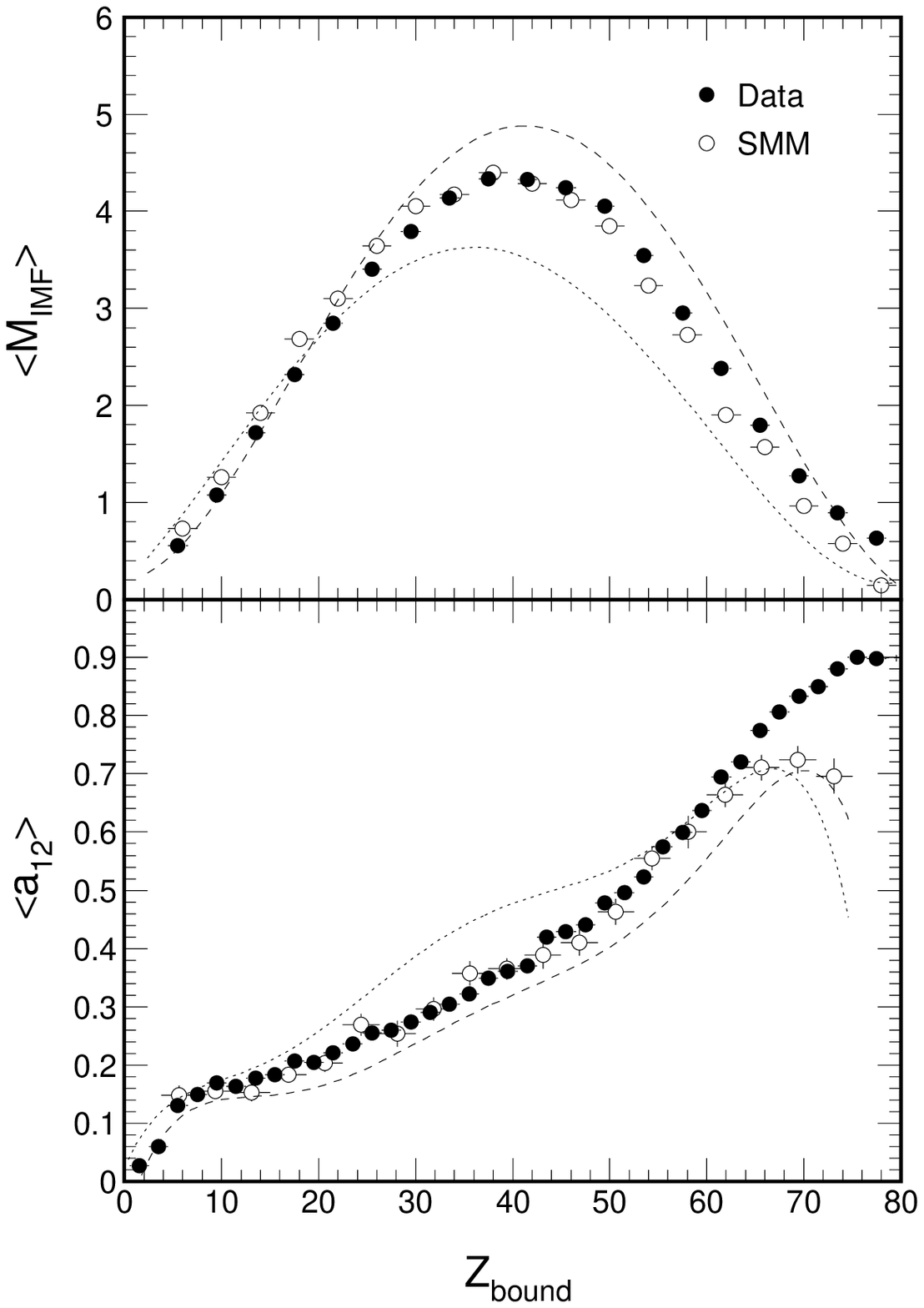,height=10cm}}
\caption[]{\it\small
Mean multiplicity of intermediate-mass
fragments $\langle M_{IMF} \rangle$ (top)
and mean charge asymmetry $\langle a_{12}\rangle$ (bottom)
as a function of $Z_{bound}$, as obtained from the calculations with the
statistical multifragmentation model (open circles), in comparison to the
experimental data for $^{197}$Au~on~$^{197}$Au at $E/A$ = 1000 MeV  
(dots, from Ref. \cite{schuetti}). The dashed and dotted lines
show the results of the calculations with excitation energies
$E_x/A$ 15\% above and 15\% below the adopted values, respectively.
Note that the trigger threshold affected the data of Ref. \cite{schuetti}
at $Z_{bound} \ge$ 65 (from Ref. \cite{hongfei}).
}
\end{figure}

The success of the statistical multifragmentation model in describing
the observed population of the partition space may be seen as a
further argument for equilibration.
Here the main task consists of finding an appropriate ensemble of
excited nuclei to be subjected to the multi-fragment decay according to the 
model prescription. Starting from the entrance channel may not
necessarily provide sufficiently
realistic ensembles, even though a good description of the
fragment correlations was obtained with the 
quantum-molecular-dynamics model coupled
to the statistical multifragmentation model \cite{konop93}.
An alternative method consists of using empirically derived ensembles.
Near perfect
descriptions of the measured correlations, including their dispersions
around the mean behaviour, can be achieved \cite{botv95,hongfei}.
The mathematical procedure of backtracing allows for
studying the uniqueness of the obtained solutions and their
sensitivities to the observables that were used to generate 
it \cite{deses96}.

The ensemble derived empirically for the reaction $^{197}$Au on $^{197}$Au 
at 1000 MeV per nucleon is shown in Fig. 2. Its capability of 
reproducing the measured mean multiplicity of intermediate-mass
fragments and the mean charge asymmetry of the two heaviest fragments
is illustrated in Fig. 3 where the dashed and dotted lines show the
model results for $E_x/A$ chosen 15\% above and below the adopted values.
In the region $Z_{bound} >$ 30, the mean excitation energy of the 
ensemble of spectator nuclei was found to be well
constrained by the mean fragment multiplicity alone.
At $Z_{bound} \approx$ 30 and below, the charge asymmetry
was a necessary second constraint while, at
the lowest values of $Z_{bound}$, neither the multiplicity nor
the asymmetry provided rigid constraints on the excitation energy.

The spectator source, well localized in rapidity \cite{schuetti} 
and, apparently, exhibiting so many signs of equilibration, seems an
excellent candidate for studying the nuclear phase diagram. Dynamical 
studies also support this conclusion \cite{fuchs97,goss97}.
There are
limitations, however, which are mainly seen in the emission of nucleons and 
very light particles. Here the components from reaction stages that lead 
to the formation of the spectator system and from its subsequent breakup 
overlap, thereby creating difficulties for the extraction of the
equilibrium properties.
This is particularly apparent in the case of the excitation energy 
which will be discussed in the next section.
\vspace{0.8cm}

\noindent
{\large\bf EXCITATION ENERGY}
\vspace{0.3cm}

A method to determine the excitation energy from the experimental data
was first presented by Campi {\it et al.} \cite{campi94}
and applied to the earlier $^{197}$Au + Cu data \cite{kreutz}.
It is based on the idea of calorimetry which
requires a complete knowledge of all decay products, including their
atomic numbers, masses, and kinetic energies. In this work, the 
measured abundances for $Z \ge$ 2 were used and, e.g., the yields of 
hydrogen isotopes were deduced by extrapolating to $Z$ = 1.
In the same type of analysis with the more recent data
for $^{197}$Au + $^{197}$Au at 600 MeV per nucleon, the data on neutron
production measured with LAND were taken into 
account \cite{poch95}.
Since the hydrogen isotopes were not detected
assumptions concerning the overall $N/Z$ ratio of the spectator,
the intensity ratio of protons, deuterons, and tritons, and the
kinetic energies of hydrogen isotopes had to be made.

\begin{figure}[ttb]
   \centerline{\epsfig{file=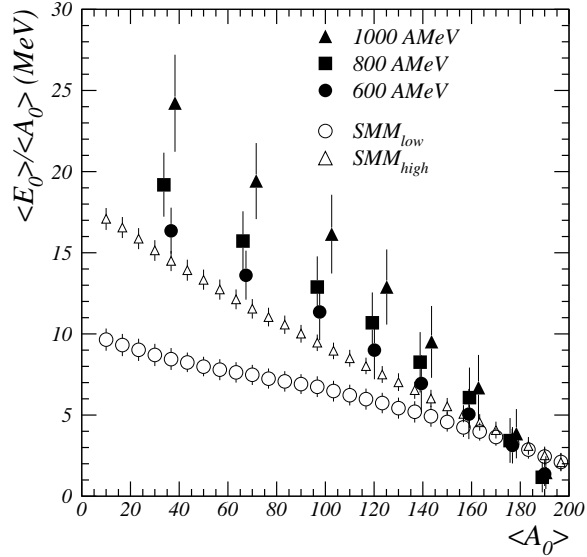,height=8cm}}
\caption[]{\it\small
Mean excitation energy per nucleon 
$\langle E_0 \rangle/\langle A_0 \rangle$ as a function of the spectator
mass $\langle A_0 \rangle$ for the reaction $^{197}$Au on $^{197}$Au 
at three bombarding energies as indicated.
The open symbols denote the upper and lower limit of the mean
excitation energy that may be derived from analyses with the 
statistical multifragmentation model (from Ref. \cite{gross}). 
}
\end{figure}

The latest evaluation of the excitation energy included 
the measured neutron data for three bombarding energies and the data 
for hydrogen emission from the target spectator at 1000 MeV per
nucleon \cite{gross}. 
For the case of 600 MeV per nucleon, the difference to the published energy 
values \cite{poch95} amounted to about 10\%.
More importantly, however, the deduced spectator energy was found 
to depend considerably on 
the bombarding energy (Fig. 4). It increases by about 30\% 
over the range 600 to 1000 MeV per nucleon
which is in contrast to the universality observed for
other observables (previous section). 
The origin of this rise lies solely in the behavior of the mean
kinetic energies of neutrons in the spectator frame (Fig. 5).
They have a large effect on the deduced total excitation
energy, first, because the neutron multiplicity is large and, second,
because the hydrogen isotopes, measured only at 1000 MeV per nucleon,
were assumed to scale in the same way as the neutrons do. 

\begin{figure}[ttb]
   \centerline{\epsfig{file=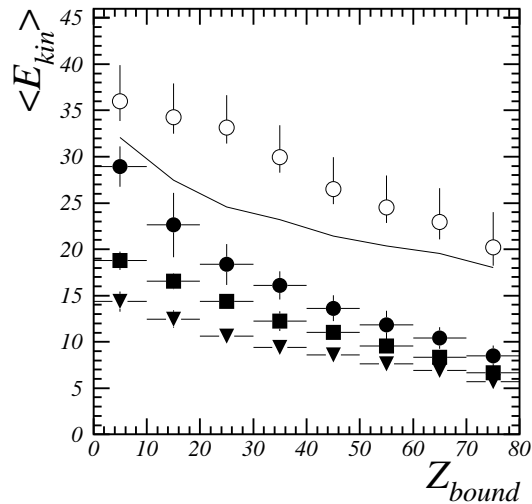,height=7cm}}
\caption[]{\it\small
Mean kinetic energy of neutrons (full symbols)
in the rest frame of the projectile
spectator as obtained in measurements including LAND for 
the reaction $^{197}$Au on $^{197}$Au at $E/A$ = 600 (triangles), 
800 (squares), and 1000 MeV (circles).
For $E/A$ = 1000 MeV, the
measured mean kinetic energies of protons (open circles)
are shown in comparison to the sum of the neutron kinetic energies and 
an estimated Coulomb contribution (full line, from Ref. \cite{gross}).
}
\end{figure}

It is obvious that this uncertainty represents a considerable problem,
a memory of the entrance channel is inconsistent with the idea of 
measuring thermodynamic properties of an equilibrated breakup
state. It is reasonable to assume that part of
the experimentally determined energy may be due to 
pre-equilibrium or pre-breakup emission, even though the analysis
of nucleon emission is restricted to the data at forward (backward)
angles in the projectile (target) frame (Fig. 6).
The experimental excitation 
energies are larger than the range of energies potentially consistent with 
the statistical multifragmentation model (Fig. 4), and the spectra of
hydrogen isotopes, measured with the high-resolution telescopes,
exhibit yields and slopes much larger than predicted by the 
model \cite{hongfei}.
The process of spectator formation involves
secondary scatterings of fireball nucleons on spectator matter
which may generate a
pre-breakup source centered close to the spectator rapidity.
Experimentally, the next step should consist of complementing 
the neutron data with equivalent data for proton and 
light-charged-particle emission at several bombarding energies. 
\vspace{0.8cm}

\begin{figure}[ttb]
   \centerline{\epsfig{file=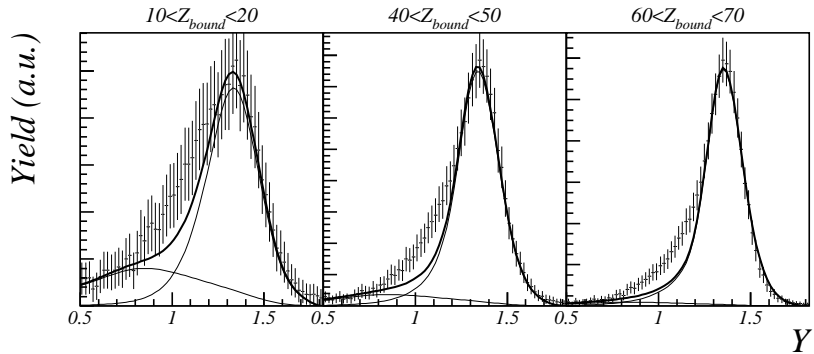,width=0.8\textwidth}}
\caption[]{\it\small
Rapidity spectra of neutrons measured with LAND for three bins in
$Z_{bound}$ for the reaction $^{197}$Au on $^{197}$Au at $E/A$ = 1000 MeV
and a cut in transverse momentum of 35 MeV/c $\le p_t \le$ 70 MeV/c.
The thick lines represent the result of fits assuming a projectile and a
mid-rapidity source. The individual contributions of these sources
are given by the thin lines (from Ref. \cite{gross}).
}
\end{figure}

\noindent
{\large\bf TEMPERATURE}
\vspace{0.3cm}

The shape of the caloric curve \cite{poch95}, reminiscent of
first-order phase transitions in ordinary liquids, and its similarity
to predictions of microscopic statistical 
models \cite{gross90,bond95,hongfei},
has initiated a widespread discussion of whether nuclear temperatures
of this magnitude can be measured reliably
(see Refs. \cite{xi96,gulm97}
and references given in these recent papers)
and whether this observation may indeed be linked to
a transition towards the vapor phase \cite{natowitz,more96}.

Here we restrict ourselves to new results obtained from the study of the 
target spectator at backward angles in the laboratory 
in $^{197}$Au + $^{197}$Au collisions at 1000 MeV per nucleon.
From the isotopically resolved yields of hydrogen, helium, and lithium 
isotopes breakup temperatures $T_{{\rm HeLi}}$, $T_{{\rm Hepd}}$,
and $T_{{\rm Hedt}}$ were derived \cite{albergo}.
The corrections for
sequential feeding of the ground-state yields, based on calculations
with the quantum statistical model \cite{konop94},
resulted in good qualitative agreement for the three temperature 
observables \cite{hongfei}.

\begin{figure}[ttb]
   \centerline{\epsfig{file=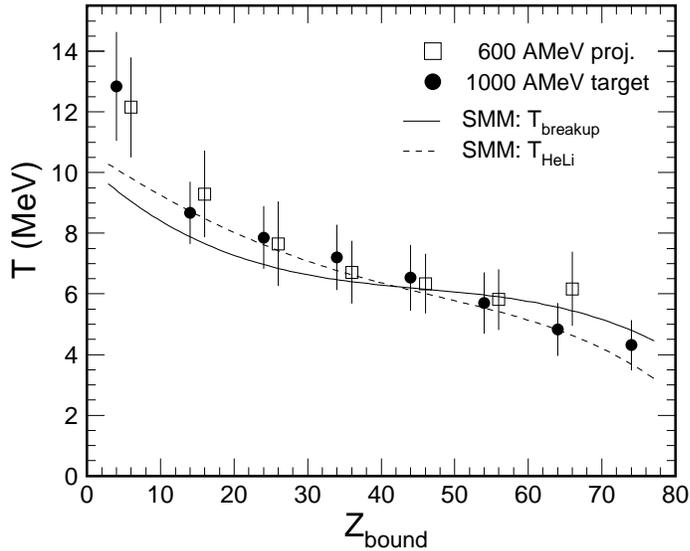,height=8cm}}
\caption[]{\it\small
Temperatures $T_{{\rm HeLi}}$ of the target spectator for
the reaction $^{197}$Au on $^{197}$Au at $E/A$ = 1000 MeV (dots)
and of the projectile spectator
at $E/A$ = 600 MeV (open squares)
as a function of $Z_{bound}$. The data symbols represent averages over
bins of 10-units width and, for clarity, are laterally displaced by 1 unit
of $Z_{bound}$. Statistical and systematic contributions are included in
the displayed errors.
The lines are smoothed fit curves describing
the breakup temperature $T_{breakup}$ (full line) and the
isotopic temperature $T_{{\rm HeLi}}$ (dashed line) calculated with the
statistical multifragmentation model (from Ref. \cite{hongfei}).
}
\end{figure}

The corrected temperature $T_{{\rm HeLi}}$ is shown in Fig. 7.
With decreasing $Z_{bound}$, it increases
from $T$ = 4 MeV for peripheral collisions to about 10 MeV
for the most central collisions.
Within the errors, these values 
are in good agreement with those measured
for projectile spectators
in the same reaction at 600 MeV per nucleon. 
In both cases, the displayed data symbols represent the mean values of the 
range of
systematic uncertainties associated with the two different experiments
while the errors include both statistical
and systematic contributions. The
projectile temperatures are the result of a new analysis of the original
data and are somewhat higher, between 10\% and 20\%,
than those reported previously \cite{poch95}.
Their larger errors follow from a
reassessment of the potential $^4$He contamination of the $^6$Li yield
caused by $Z$ misidentification.
The invariance of the breakup temperature
with the bombarding energy is consistent with
the observed universality of the spectator decay.

Calculations with the statistical multifragmentation model were performed
for the ensemble of excited spectator nuclei shown in Fig. 2.
Results have already been shown in Fig. 3.
The solid line in Fig. 7 represents the thermodynamic
temperature $T$ obtained in these calculations.
With decreasing $Z_{bound}$, it increases monotonically from
about 5 to 9 MeV. Over a wide range of $Z_{bound}$
it remains close to $T$ = 6 MeV which
reflects the plateau predicted by the statistical multifragmentation model
for the range of excitation energies
3 MeV $\le E_x/A \le$ 10 MeV \cite{bond95}.
In model calculations performed for a fixed spectator mass, the plateau
is associated with a strong and monotonic rise of the fragment 
multiplicities.
Experimentally, due to the decrease of the spectator mass with
increasing excitation energy (Fig. 2), the production of intermediate-mass
fragments passes through a maximum in the corresponding
range of $Z_{bound}$ from about 20 to 60 (Fig. 3).

The dashed line gives the temperature $T_{{\rm HeLi}}$ obtained from
the calculated isotope yields. Because of sequential feeding, it
differs from the thermodynamic temperature, the uncorrected
temperature $T_{{\rm HeLi},0}$ being somewhat lower.
Here, in order to permit the direct comparison
with the experimental data in one figure,
we display $T_{{\rm HeLi}}$ which has been
corrected in the same way with the factor 1.2
suggested by the quantum statistical model. The calculated $T_{{\rm HeLi}}$
exhibits a more continuous rise with decreasing $Z_{bound}$
than the thermodynamic temperature and is in
very good agreement with the measured values.
We thus find that, with the parameters needed to reproduce
the observed charge partitions, this temperature-sensitive observable
is well reproduced. A necessary requirement
for a consistent statistical description of the spectator fragmentation
is thus fulfilled.

\begin{figure}[ttb]
   \centerline{\epsfig{file=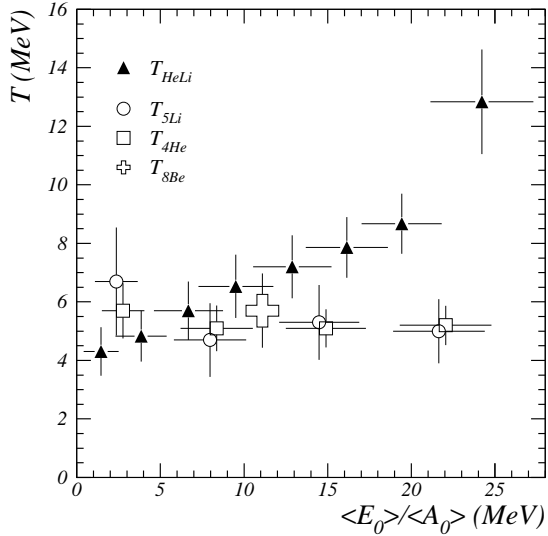,height=8cm}}
\caption[]{\it\small
Measured isotope temperature $T_{{\rm HeLi}}$ (full triangles) 
and excited-state temperatures (open symbols) as a function of the 
experimental excitation energy per nucleon
$\langle E_0 \rangle/\langle A_0 \rangle$ for the reaction 
$^{197}$Au on $^{197}$Au at $E/A$ = 1000 MeV. The
indicated uncertainties are mainly of systematic origin. The displayed
value of $T_{{\rm 8Be}}$ represents an average over the full range
of $\langle E_0 \rangle/\langle A_0 \rangle$ (from Ref. \cite{gross}).
}
\end{figure}

In the same experiment excited-state 
temperatures \cite{poch87,kunde91}
were determined from the populations of 
particle-unstable resonances measured with the Si-CsI hodoscopes.
The peak structures were identified by using the technique of correlation
functions, and background corrections were based on results obtained for
resonance-free pairs of fragments with $Z \le$ 3,
such as p-d, d-d, up to $^3$He-$^7$Li. 
Correlated yields of p-t, p-$^4$He, d-$^3$He, $^4$He-$^4$He, 
and p-$^7$Li coincidences and $^4$He singles yields were 
used to deduce temperatures from the
populations of states in $^4$He (g.s.; group of three states at
20.21 MeV and higher), $^5$Li (g.s.; 16.66 MeV),
and $^8$Be (3.04 MeV; group of five states
at 17.64 MeV and higher).
The probabilities for the coincident detection
of the decay products of these resonances were calculated with a
Monte-Carlo model \cite{kunde91,serf97}. The
uncertainty of the background subtraction is the main
contribution to the errors of the deduced temperatures.

The values for the three excited-state temperatures are given 
in Fig. 8 as a function of the experimental excitation energy
$\langle E_0 \rangle/\langle A_0 \rangle$.
Mutually consistent with each other, they appear to be virtually 
independent of the excitation energy, centering about a mean value
of $\approx$ 5 MeV. This is in striking contrast to the 
monotonically rising isotope temperature $T_{{\rm HeLi}}$ 
which is shown in comparison.

A saturation of excited-state temperatures and a similar 
difference to the behavior of isotope temperatures has also 
been observed in central $^{197}$Au + $^{197}$Au collisions
at incident energies $E/A$ = 50 MeV to 200 MeV \cite{serf98}.
The interpretation given there starts from the fact that
the excited states used for the temperature evaluation are very specific
quantum states which may not 
exist in the nuclear medium in identical 
forms \cite{dani92,roepke,alm95}.
The observed asymptotic states can develop and 
survive only at very low densities that may not be reached before
the cluster is emitted into vacuum.
Accordingly, the excited-state populations should reflect the
temperature and its fluctuations at this final stage of fragment emission.
The obtained mean value near 5 MeV is not inconsistent with results of 
dynamical calculations based on the BUU model \cite{fuchs97}.
\vspace{0.8cm}

\noindent
{\large\bf DENSITY}
\vspace{0.3cm}

Expansion is a basic conceptual feature of both the statistical 
multifragmentation and the liquid-gas phase transition.
A volume of about six to eight times that occupied at saturation density
is assumed in the statistical multifragmentation models while the critical
volume in the case of infinite nuclear matter has about three times 
the saturation value. The experimental confirmation of expansion 
or low breakup density is therefore of the highest significance.

In central collisions of heavy nuclei, expansion is evident from the
observation of radial collective 
flow \cite{reis97a}.
Significant radial flow is not observed in spectator decays, and 
evidence for expansion has been obtained, indirectly, from model 
comparisons.
Models that assume sequential emission from the 
surfaces of nuclear systems at saturation density underpredict 
the fragment multiplicities while those assuming expanded breakup
volumes yield satisfactory 
descriptions of the populated partition space \cite{hubel92}.
The disappearance of the Coulomb peaks in the kinetic-energy spectra of 
emitted light particles and fragments, associated with increasing fragment 
production, provides additional evidence consistent with volume 
emission or emission from expanded systems \cite{milkau91}.

\begin{figure}[ttb]
   \centerline{\epsfig{file=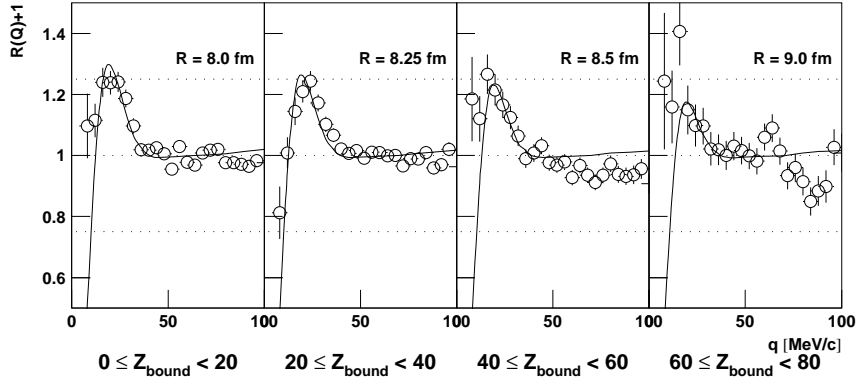,width=0.8\textwidth}}
\caption[]{\it\small
Proton-proton correlation functions for the indicated intervals 
of $Z_{bound}$, measured at $\Theta_{lab} \approx 135^{\circ}$ for the 
reaction $^{197}$Au on $^{197}$Au at $E/A$ = 1000 MeV. The full lines 
are obtained from the analysis with the Koonin-Pratt formalism, the resulting
source radii are indicated (from Ref. \cite{fritz}).
}
\end{figure}

Interferometric methods permit experimental determinations of the breakup
volume or, more precisely, of the space-time extension of where the 
emitted products had suffered their last collision \cite{ardouin}. 
In the present case of spectator decay at relativistic energies
the time scales should be rather 
short and the data, in good approximation, may be directly related to the
breakup volume that is of interest here. 

Proton-proton correlation 
functions measured at angles of $\Theta_{lab} \approx$ 135$^{\circ}$ 
are shown in Fig. 9. They are characterized by a depression 
at small relative momenta, caused by Coulomb repulsion, and by 
a peak near $q$ = 20 MeV/c that is caused by the S-wave nuclear 
interaction. Its comparatively small amplitude signals a large spatial
extension of the proton source. The quantitative analysis of these data
was performed with the Koonin-Pratt formalism \cite{pratt}.
A uniform sphere was assumed for the
proton source. The deduced radii are of the order of 8 to 9 fm which
is distinctly larger than the radius of 6.5 to 7 fm 
of a gold nucleus at normal density. 
The structure of the correlation functions is somewhat obscured at 
larger $Z_{bound}$ but there is no indication that the radii and thus the 
volume should
significantly change with impact parameter. 
The derived density, however, decreases considerably 
with increasing centrality, 
caused by the changing number of spectator constituents $A_0$ (Fig. 10). 
These spectator masses result from the calorimetric analysis described 
above and are found to be in good agreement with the prediction 
of the geometric participant-spectator model \cite{gosset}. 
The mean
relative density decreases to values below $\rho /\rho_0$ = 0.2 for the
most central bin, i.e. smallest $Z_{bound}$. The deduced values as well
as the variation with centrality compare well with the densities
entering the statistical multifragmentation model in the version
that uses a fixed cracking distance for the placement of fragments inside 
the breakup volume \cite{bond95}.
We recall here that the proton multiplicities and kinetic-energy spectra
indicate a predominantly pre-breakup emission. This does not necessarily 
exclude the possibility that their interaction with the forming 
spectator matter causes the interferometric picture to reflect the 
extension of the latter.

Besides the proton-proton correlations also correlations of other light 
charged particles were used to determine breakup radii. Pronounced 
resonances are exhibited by the p-$\alpha$ ($^5$Li, g.s.), 
d-$\alpha$ ($^6$Li, 2.19 MeV), and t-$\alpha$ ($^7$Li, 4.63 MeV) 
correlation functions. Their peak heights were analyzed using the
numerical results of Boal and Shillcock \cite{boal}.
The deduced values, within errors are in qualitative agreement 
with the proton values.
Their much smaller error bars demonstrate that higher 
accuracies may be reached with these pronounced
resonances of particle unbound states in light nuclei.
The further development of the formalism needed for their quantitative
interpretation seems therefore highly desirable.
\vspace{0.8cm}

\begin{figure}[ttb]
   \centerline{\epsfig{file=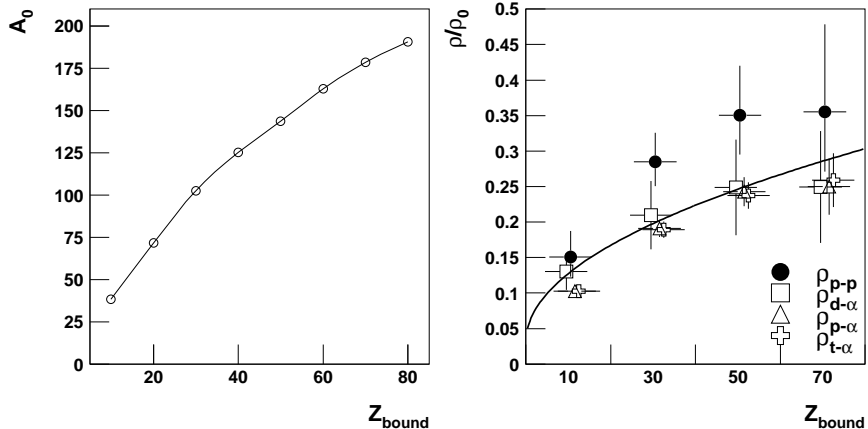,width=0.8\textwidth}}
\caption[]{\it\small
Spectator mass $\langle A_0 \rangle$ (left panel)
and relative density $\rho /\rho_0$ (right panel) 
as a function of $Z_{bound}$ for the reaction
$^{197}$Au on $^{197}$Au at $E/A$ = 1000 MeV. The lines are meant to guide 
the eye. The densities were derived from proton-proton correlations
(Fig. 9) and from correlation functions constructed for light charged 
particles as indicated (open symbols, from Ref. \cite{fritz}).
}
\end{figure}

\noindent
{\large\bf SUMMARY AND PERSPECTIVES}
\vspace{0.3cm}

New results for the mass, excitation energy, temperature, and density
of excited spectator systems at breakup have been presented. The discussion 
of these data was meant to demonstrate that methods exist to determine 
these thermodynamic variables from the experiment. It was also intended
to show that they are not without problems and that, perhaps, serious 
conceptual difficulties exist which may make it very difficult to 
arrive at unambiguous results. The problem caused for the excitation energy 
by the dynamical process of spectator formation is an example.

It should be rewarding to further pursue this program, 
not only because of the hope to better identify signals of the 
liquid-gas phase transition but also 
because unexpected results may appear, as demonstrated for
the temperatures. The saturation
of the excited-state temperatures, according to the interpretation given 
here, is an interesting in-medium effect in itself. It also has
the consequence that it may provide us  
with a means to determine rather reliably the internal temperatures of 
fragments at their final separation from the system.

The interferometry with light charged particles confirms the low density
of the breakup configuration. For more precise evaluations the
formalisms needed to deduce radii and densities from 
interferometric measurements will need continuing development.
Measurements of radii for different fragment species, possibly emitted at 
different stages of the reaction, may allow us to test and refine the 
otherwise so successful picture of the single breakup state in
the spectator decay.

\end{document}